\documentclass{PoS}

\newcommand*{\La}{\cal{L}}
\newcommand*{\no}{\noindent}
\newcommand*{\bea}{\begin{eqnarray}}
\newcommand*{\eea}{\end{eqnarray}}
\newcommand*{\be}{\begin{equation}}
\newcommand*{\ee}{\end{equation}}
\newcommand*{\pd}{\partial}
\newcommand*{\pdm}{\pd_{\mu}}
\newcommand*{\pdn}{\pd_{\nu}}

\newcommand*{\pref}[1]{(\ref{#1})}

\newcommand*{\mn}{{\mu\nu}}

\newcommand*{\nn}{\nonumber}

\title{On gauge fixing}

\ShortTitle{On gauge fixing}

\author{\speaker{Axel Maas}\thanks{Supported by the FWF under grant number M1099-N16.}\\
        Karl-Franzens-University Graz, Universit\"atsplatz 5, A-8010 Graz, Austria\\
        E-mail: \email{axelmaas@web.de}}

\abstract{Gauge fixing is a useful tool to simplify calculations. It is also valuable to combine different methods, in particular lattice and continuum methods. However, beyond perturbation theory the Gribov-Singer ambiguity requires further gauge conditions for a well-defined gauge-fixing prescription. Different additional conditions can, in principle, lead to different results for gauge-dependent correlation functions, as will be discussed for the example of Landau gauge. Also the relation of lattice and continuum gauge fixing beyond perturbation theory will be briefly outlined.}

\FullConference{The XXVIII International Symposium on Lattice Field Theory, Lattice2010\\
		June 14-19, 2010\\
		Villasimius, Italy}

\begin{document}

\section{Why gauge fixing?}

Gauge fixing is essentially a choice of coordinates in the field configuration space, and thus nothing else than choosing a suitable coordinate system for calculations. In fact, to calculate observables, it is intrinsically not necessary to fix a gauge. However, as often, it is much more simpler to select a useful coordinate system than not. Hence, almost all current calculations, from perturbation theory to lattice calculations of the hadron spectrum, use at one time or another gauge-fixing.

One application of gauge fixing has been calculations beyond perturbation theory when combining different methods. A particular useful example is the the combination of lattice and continuum functional methods \cite{Fischer:2008uz}. In this case, the lattice calculations are providing gauge-fixed correlation functions in the domain where they are rather reliable. Functional methods then extend these calculations to regions where lattice calculations are not feasible, like finite density, the chiral limit, disparate energy scales, and so on.

Furthermore, all non-perturbative methods require at some point systematically uncontrolled approximations. For functional methods, these are truncations. For lattice calculations, these are finite volume and discretization, for which no analytical tools are available, and it is necessary to refer to extrapolations. Thus a comparison of different non-perturbative methods helps to control these systematic errors, in particular when it comes to predictions for experiments. Comparing then the simplest, and thus in a gauge theory gauge-dependent, correlation functions helps to establish systematic coherence. However, gauge fixing becomes itself a problem beyond perturbation theory.

\section{Gauge fixing as a selection process}

The original idea of a gauge theory is to trade a redundant set of coordinates for locality. As a consequence, it is possible to describe, e.\ g., Yang-Mills theory using only local interactions\footnote{It is in principle possible to go back to a formulation without redundant coordinates, like in the loop-formalism of lattice gauge theory. However, the variables and interactions then become inherently non-local.}. The price paid is that not every field configuration describes a different physical process. In continuum gauge theories, there is an denumerable infinite degeneracy in terms of physical observables for the local variables, the gauge fields: Each physical situation is represented by a continuous gauge orbit of gauge-equivalent fields, i.\ e., field configurations which can be transformed into each other by a gauge transformation. In the case of Yang-Mills theory
\bea
\La&=&-\frac{1}{4}F_\mn^a F^\mn_a\nn\\
F_\mn^a&=&\pdm A_\nu^a-\pdn A_\mu^a+g f^a_{bc} A_\mu^b A_\nu^c\nn
\eea
\no an infinitesimal move along this gauge orbit is given by
\bea
A_\mu^a&\to& A_\mu^a+D_\mu^{ab}\phi_b\nn\\
D_\mu^{ab}&=&\delta^{ab}\pdm+g f^{ab}_c A_\mu^c\nn.
\eea
\no with $A_\mu^a$ the gauge fields, $g$ the gauge coupling, $f^{abc}$ the structure constants of the associated gauge algebra, and the $\phi^a$ are arbitrary functions.

When performing a path integral quantization, the problem is how to ensure that each gauge orbit is counted the same number of times. This can be ensured by either of two ways \cite{Bohm:2001yx}. One way is that an average over all possible representatives of the gauge orbit is performed, which is the basic idea of gauges like the Feynman gauge. The second possibility is to select a uniquely characterized representative for each gauge orbit. This is done, e.\ g., in Landau gauge, which chooses perturbatively the one representative which satisfies
\be
\pd^\mu A_\mu^a=0\label{lg}
\ee
\no These prescriptions can be implemented in a path integral by inserting an appropriate weight function or a $\delta$-function, respectively \cite{Bohm:2001yx}. In the continuum, such prescriptions are usually not evaluated as a selection procedure on Gribov copies, but are rewritten with the help of auxiliary fields, the ghost fields. Lattice calculations, on the other hand, have the advantage to be able to really select gauge copies for the calculation of the path integral. Thus, in lattice calculations a literal implementation of a gauge condition is possible.

\section{The Gribov-Singer ambiguity}

The problem of gauge fixing beyond perturbation theory is that it becomes complicated to specify gauge conditions \cite{Gribov:1977wm}. The reason is that the gauge algebras of Yang-Mills theory are simple Lie algebras \cite{Singer:1978dk}. As a consequence there is no single coordinate system which is able to cover the whole field configuration space. Hence, it is not possible to give a local (and thus single coordinate system) prescription to characterize the selection criterion for representatives of the gauge orbits. This problem is known in the context of covariant gauges as the Gribov-Singer ambiguity, but it occurs in one disguise or the other for all gauges studied so far, as long as a continuum formulation is desired. In case of the Landau gauge this problem boils down to the fact that beyond perturbation theory there is more than one solution to the equation \pref{lg}, and it is not possible to impose purely local constraints to single out one and only one representative for each gauge orbit.

However, it should be remarked that there is no conceptual difference between Gribov copies, i.\ e.\ the multiple solutions of \pref{lg}, and ordinary gauge copies. All of them are just regular representatives of a gauge orbit. The only difference is that two Gribov copies are not infinitesimally close to each other, as otherwise a local distinction would be possible. They are therefore separated by non-infinitesimal gauge transformations. Still, they are just ordinary gauge copies, and their removal is in no way different from the removal of other gauge copies. It is just that because of the geometrical structure of the gauge orbits it is not possible to do so by local conditions.

It is of course possible, instead of trying to specify with non-local gauge condition such a single representative, to just enlarge the averaging procedure of perturbation theory such as to encompass also the Gribov copies. This case has additional challenges, known as the Neuberger problem \cite{Neuberger:1986xz}. However, these can be overcome \cite{smekal}, and such gauges can indeed be constructed.

Irrespective of which possibility is chosen there is no reason to suspect a-priori that there exists a unique way of how to extend the selection process. To discuss this in more detail, the setting will be simplified by first reduce the gauge orbits to Landau gauge i.\ e., all remaining representatives of the gauge orbit, the residual gauge orbit \cite{Maas:2008ri}, fulfill \pref{lg}.

\section{The example of Landau gauge(s)}

The residual gauge orbits can then be partitioned in Gribov regions \cite{Gribov:1977wm}, which give the orbit space a shell-structure, with the shell surrounding the origin also including perturbation theory. It can then be shown that each shell contains at least one representative of each gauge orbit, but that there exists also for at least some gauge orbits multiple representatives in each shell \cite{Zwanziger:2003cf}. Thus, each shell contains all physical information. The shells are called Gribov regions, their boundaries are called Gribov horizons.

Among all possibilities how to treat the residual gauge orbits, two have received most attention.

One averages over all Gribov regions \cite{Neuberger:1986xz}. This construction can be shown to harbor a full non-perturbatively well-defined version of BRST symmetry \cite{Neuberger:1986xz,smekal}, which has the same algebraic structure as the perturbative one. However, in any calculation this requires to take into account the signed measure of each Gribov copy, which so far cannot be turned into a simple expression.

The second one reduces the residual gauge orbit to the innermost Gribov region. It then remains to specify further how to deal with the remaining representatives. Three different, equally valid, choices have been especially pursued in lattice gauge theory recently:
\begin{itemize}
 \item Minimal Landau gauge \cite{Maas:2008ri,Cucchieri:1995pn,Maas:2009se}. In this case for each orbit a random representative is chosen among the possible ones. Results indicate that for correlation functions this may be equivalent to average over all representatives of each residual gauge orbit.
 \item Absolute Landau gauge \cite{Maas:2008ri,Zwanziger:2003cf,Bornyakov:2009ug}. In this case the representative is selected which minimizes a certain non-local functional (the integral of the trace of the gluon propagator \cite{Maas:2008ri}) absolutely, though this has still a minor problem with topological identifications of certain gauge copies. It has been conjectured that for correlation functions made from a finite polynominal of the gauge fields this should yield the same correlation function as the minimal Landau gauge \cite{Zwanziger:2003cf}, which is supported by available results \cite{Bornyakov:2009ug}.
 \item Landau-$B$ gauges \cite{Maas:2009se}. In this case a representative is chosen in which a certain non-local quantity (the ghost propagator at zero momentum) is agreeing best with a predefined value (called $B$). These gauges have shown the largest variability of correlation functions yet, though they tend to agree for finite polynomials of the fields also with the minimal Landau gauge.
\end{itemize}
All of them provide a well-defined prescription how to select a representative. The minimal Landau gauge depends of course on how well the random choice is implemented such that all representatives are equally sampled. Though common algorithms seem to perform rather well \cite{Maas:2009se}, there is no proof. Fixing the absolute Landau gauge is an NP-hard problem of spin-glass type, and therefore no guarantee exists whether any existing algorithm attempting to fix to this gauge actually does this, though the results seem to support it \cite{Maas:2008ri,Bornyakov:2009ug}. Finally, the Landau-$B$ gauge has not been shown to be able to differentiate between Gribov copies perfectly, i.\ e.\ that each Gribov copy has a different $B$ value. However, it just averages in the sense of minimal Landau gauge over remaining copies \cite{Maas:2009se}, which finally makes up for a well-defined prescription in the same sense as minimal Landau gauge.

\section{Scaling {\underline{and}} decoupling or why there could be place for more than one solution}

To put these constructions into perspective, it is necessary to make a short detour.

It is a remarkable fact that both the Dyson-Schwinger equations (DSEs) and the functional renormalization group equations (FRGs) are form-equivalent, irrespective of whether they are formulated within a single Gribov region or not \cite{Zwanziger:2001kw}. Therefore, the solution manifold of these equations has to contain not only the solution inside the first Gribov region, but also in every other Gribov region, and in the whole of the system without specifying the Gribov region. It is furthermore quite important that the equations are not closing on themselves alone, but at least one external specification has to be introduced, e.\ g., the ghost dressing function at zero momentum, at least for the truncations where this has been investigated so far \cite{Fischer:2008uz,Boucaud:2008ky}. It is furthermore intriguing that the situation in Coulomb gauge is quite similar \cite{Watson:2010cn}.

Using this one parameter, and imposing furthermore that the ghost dressing function should be positive (semi-)definite, an one-parameter family of solutions for these equations are found \cite{Fischer:2008uz,Boucaud:2008ky}. One of the end-points of the permitted interval is special, as it corresponds to a qualitatively different kind of solution, the so-called scaling solution, which is characterized by critical behavior in the infrared \cite{Fischer:2008uz}. The remaining solutions are characterized by a screening behavior, and therefore the degrees of freedom largely decouple, except for the photon-like ghost.

The natural questions arising are: Assuming that the existence of these solutions is not a pure truncation artifact, to which of the Gribov regions do they belong? And is it possible to reproduce them on the lattice? At least for one of the decoupling-type solutions the latter question can be answered by yes. It is found equally well in one particular Landau-$B$ gauge \cite{Maas:2009se} and in the minimal Landau gauge \cite{Fischer:2008uz,Cucchieri:2007rg,Bogolubsky:2009dc}. Also, the absolute Landau gauge appears to show a decoupling behavior \cite{Bornyakov:2009ug}, in contrast to the original expectations \cite{Maas:2008ri,Zwanziger:2003cf}. However, it is not yet clear, whether it produces the same decoupling solution in the infinite volume and continuum limit \cite{Maas:2008ri}.

The interesting question is: Can the other ones be reproduced on the lattice? Given the current algorithms, the necessary precondition for a positive answer is, whether they belong to the first Gribov region. Since outside the first Gribov region, the Faddeev-Popov operator acquires more and more negative eigenvalues, it appears unlikely that the ghost propagator can maintain a single sign. In 1+1-dimensional Coulomb gauge there is actually a proof that the ghost dressing function is only positive inside the first Gribov region \cite{Reinhardt:2008ij}. This motivates that the whole family of solutions could be obtainable inside the first Gribov region, and thus with current gauge-fixing algorithms.

However, if they are there, then they have necessarily to correspond trivially to a different selection of Gribov copies than the minimal Landau gauge, because they are different. Thus, if they can be found, this implies that they are the solutions obtained in different non-perturbative realizations of the Landau gauge. At the current time, such a one-parameter-family is found using Landau-$B$ gauges, but this is only obtained for rather small lattices and coarse discretizations \cite{Maas:2009se}. To study this further is mandatory, given the experience with the minimal Landau gauge \cite{Fischer:2008uz,Cucchieri:2007rg,Bogolubsky:2009dc}. But it also motivates that the family could be contained in the first Gribov region.

A particular case is the scaling version. Particular for a single reason: In favor of its existence in the continuum case it has been embedded \cite{Fischer:2008uz} in the construction of Kugo and Ojima \cite{Kugo:1979gm}, which required the introduction of a global BRST. As noted, this requires to average over all Gribov regions. Since the functional equations contain the information from all Gribov regions, there is no contradiction. However, if this is correct, and only then, for the question concerning the lattice realization with contemporary algorithms the problem arises if it is possible to select Gribov copies inside the first Gribov region such that they become equivalent to this average over all Gribov regions. Or, put it in another way, cancel the contribution of all other Gribov regions? If the answer is no, there is no possibility with current gauge-fixing algorithms to verify, or falsify, the existence of the scaling solution with lattice methods under the assumption that the embedding is correct. Otherwise, the same applies to scaling as applied to decoupling.

However, there are number of arguments in favor of this possibility. One is the positivity of the ghost dressing function, in particular in view of the 1+1-Coulomb case. The second is that scaling appears to be realized for all volumes studied so far in two dimensions \cite{Cucchieri:2007rg,Maas:2007uv}. Since the Kugo-Ojima/BRST construction is the same in two dimensions, at least there such a cancellation appears possible. There is then no a-priori reason that this should not also be possible in higher dimensions, and that it is not only obscured by the problem of finding the right Gribov copies. This is also supported by the fact that in three dimensions over some momentum range scaling is seen, before it finally turns decoupling \cite{Maas:2008ri,Cucchieri:2007rg}. If this is a problem of finding the right Gribov copies, it is clear that it becomes harder in three and even harder in four dimensions \cite{Maas:2009se}.

Unfortunately, this is by no means a guarantee that this can work out. Nonetheless, it is sufficient motivation to investigate this possibility. Since the Landau-$B$ gauges provide, by construction \cite{Maas:2009se}, the most divergent ghost propagator, required for the scaling case, they are the ideal tool for this search. And while available results increase the motivation \cite{Maas:2009se}, only further investigations will be able to make something close to a statement. Unfortunately, numerical lattice simulations are never able to verify or falsify such a question like the existence of a solution. Already the experience from solid-state physics teach us that there can always something unexpected happen with the next order of magnitude of volume or discretization. Thus, the combination with continuum and other methods to obtain a final answer is indispensable.

Thus, the question is therefore whether it is possible to impose Landau-$B$ gauges also in the continuum as a gauge condition. The formulation of the Landau-$B$ gauges on the lattice suggest that a property is imposed on the average, and in particular it is not needed to be exactly fulfilled on every gauge orbit. Such constraints can be realized using Lagrange multiplier, suggesting as a possible form for Landau-$B$ gauges for an operator ${\cal O}$ in the continuum the form
\bea
<{\cal O}>&=&\lim_{\xi\to 0}\int{\cal D}A_\mu{\cal D}c{\cal D}\bar{c} {\cal O}(A_\mu,c,\bar{c})\theta\left(-\pdm D_\mu^{ab}\right)e^{-\int d^4x {\La}_g}\exp\left(\frac{Z_BB}{V}\int d^dxd^dy\pdm^x\bar{c}^a(x)\pdm^yc^a(y)\right)\nn,
\eea
\no where ${\cal L}_g$ is the Faddeev-Popov Lagrangian, the $\theta$-function implements the restriction to the first Gribov region, $V$ is the volume, the integral multiplying $B$ is the ghost dressing function at zero momentum, and $Z_B$ ensures the desired renormalization properties \cite{Maas:2009se}. This is, of course, a non-local object, which explicitly breaks perturbative BRST. This is purely a speculative proposal, guided by the intuition of statistical mechanics. It should be taken only as an incentive for a line of thought in the future. But if such a realization of the Landau-$B$ gauge is possible, this would be a further step to connect the lattice and the continuum formulation, in particular also to finally realize an understanding of what minimal Landau gauge is in the continuum, what is currently far from obvious \cite{Fischer:2008uz,Maas:2008ri,Maas:2009se}.

\section{Outlook}

Gauge fixing is a useful tool to investigate many problems. It has already been quite useful in perturbation theory, and it is so beyond perturbation theory. However, we have only just begun to explore the possibilities of using the Gribov-Singer ambiguity to design gauges with desirable properties, and the investigations in Landau gauge will provide us with deep insights into what is the potential. In particular, it will provide us with an understanding to which extend we can deform the properties of correlation functions by mere gauge choices. The discovery of multiple solutions to the continuum equations, and the support of one of them by lattice methods, and algebraic arguments in favor of another only motivates us to understand better what is the role of gauge-fixing in this. But yet lacking analytic control, it is at the current time of significant importance to pursue every possibility found to map out the implications of non-perturbative gauge fixing.

\end{document}